\documentclass[12pt]{article}
\usepackage{amssymb}
\usepackage{amsmath}
\usepackage[dvips]{graphicx}

\title{Integrable models for quantum media\\excited by laser radiation:\\a method, physical interpretation, and examples}
\author{Vadim A. Savva\thanks{Stepanov Institute of Physics, Belarus National Academy of Sciences, Minsk, Belarus}, %
Vadim I. Zelenkov\thanks{International Sakharov Environmental University, Minsk, Belarus}}
\date{}
\begin{document}
\maketitle

\begin{abstract}A method to build various integrable models for description of coherent excitation of multilevel media by laser pulses is suggested.
Distribution functions over the energy levels of quantum systems depending on the time and frequency detuning are obtained. The
distributions follow from Schr\"odinger equation exact solutions and give the complete dynamical description of laser-excited quantum
multilevel systems. Interpretation based on the Fourier spectra of the probability amplitudes of a quantum system is presented. The spectra
are expressed in terms of orthonormal polynomials and their weight functions. Matrix elements of the dipole transitions between levels are
equal to coef{}ficients of the recurrence formula for the orthonormal polynomial system. Some examples are presented. The Kravchuk
oscillator family as an integrable model is constructed to describe the coherent excitation dynamics of multilevel resonance media. It is
based on the use of the Kravchuk orthogonal polynomials. The Kravchuk oscillator excitation dynamics is described by the binomial
distribution of energy level populations and the distribution parameter depends on excitation conditions. Two known basic models in quantum
physics -- the harmonic oscillator and two-level system are the special representatives of the Kravchuk oscillator family.
\end{abstract}

%%%%%%%%%%%%%%%%%%%
%       111       %
%%%%%%%%%%%%%%%%%%%

\section{Introduction}
In dynamical problems of quantum mechanics there exists a rather small number of integrable models. The harmonic oscillator and two-level
system are the most well-known and popular ones for the description of matter-field interaction \cite{landau1,baz2,shore3}. The first model
is the basic one for all quantum physics but it can not describe the saturation effect that is important in laser physics, resonance
nonlinear optics, and coherent population transfer where the second model is used.

In this paper we show that there are many various multilevel quantum systems such that their coherent dynamics in an electromagnetic field
can be described exactly in a closed form. A method to construct exact analytical solutions of coherent dynamical equations  and some
examples are presented. A family of multilevel quantum systems (Kravchuk oscillators) is constructed. The coherent dynamics of the Kravchuk
oscillators is described by the binomial distribution of the populations. The family contains both the harmonic oscillator and two-level
system as a limiting case and a special one.

The nonstationary problems of coherent excitation of quantum systems have become the subject of great interest in recent years
\cite{shore3}. The coherent excitation process takes place for a short time interval when the relaxation does not destroy the quantum
medium coherence induced by radiation. Coherent processes are used for the collisionless multiphoton excitation of molecules in the field
of laser radiation, for the laser isotope separation, in studies of coherent effects in resonance media  excited by ultrashort pulses, in
studies of decoherence channels and searching the ways of decoherence elimination with the goal of using these conditions in quantum
technologies and quantum computers, for laser controlled chemical reactions (femtochemistry) and in spectroscopy of selective excitated
molecules for studying intramolecular and intermolecular relaxation and characterization of highly exited molecules.

%%%%%%%%%%%%%%%%%%%
%       222       %
%%%%%%%%%%%%%%%%%%%

\section{Model of a quasi-resonance medium excited by laser radiation}
The model is a quantum system with the energy levels $E_{n}=\sum_{k=0}^n\hbar{\,}\omega_k; \quad n=0,1,\ldots,N;$ and the radiative
transitions between neighbouring levels. We take into account only the most intensive transitions. Each transition ${n{-}1}\leftrightarrow
n$ is characterized by the dipole matrix element $\mu_{n-1,{\,}n}=\mu_{01}f_n; \quad f_0\!=\!0; \quad f_1\!\!=\!\!1$. The function $f_n$ of
dipole transition moments is the basic important characteristic of the multilevel quantum system. It describes the dependence of the matrix
elements on energy, i.e. on the number $n$ of an ${n{-}1}\leftrightarrow n$ transition, and $f_n$ is normalized to the $0\leftrightarrow 1$
transition.

\begin{figure}[tbph]
\centering\includegraphics[width=0.3\textwidth]{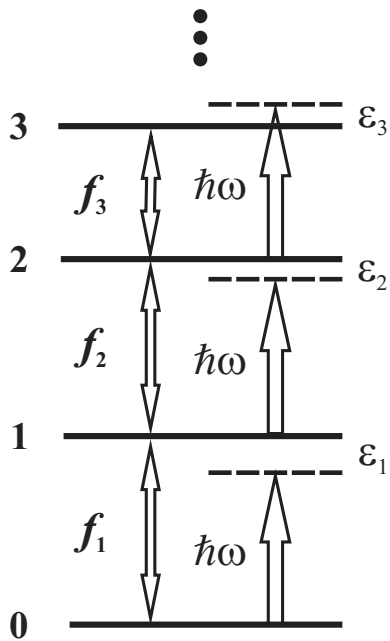}
\caption{Radiative transitions between adjacent levels; $\omega$~radiation frequency, $\varepsilon_n$~frequency detuning,
$f_n$~dipole moment functions.}
\end{figure}

It is very seldom that the level energies and transition dipole moments could be calculated from a stationary problem for a quantum system
not interacting with radiation and having the hamiltonian ${\rm \bf \hat H}_0 (z)$. As a rule, they are gained from spectral measurements.

A laser pulse  switched on at the instant $\tau=0$ has the form $\cal E(\tau)=\cal E_{\ell} \cos{} (\omega_{\ell}\,\tau)$ with the
amplitude $\cal E_{\ell}$ and carrier frequency $\omega_{\ell}$; $\tau$ is the time in seconds. The pulse causes the transition of a
molecule from initial state $E_0 =0$ to higher energy levels. The model is depicted in Figure\;1. It is a phenomenological model and it
extends the well-known two-level one \cite{landau1, allen eberly}. For the coherent dynamics of multilevel media this model has been used
in many papers including \cite{eberly shore bial1, makarov}.

%%%%%%%%%%%%%%%%%%%
%       333       %
%%%%%%%%%%%%%%%%%%%

\section{Dynamical equations and solution method}
In the rotating-wave approximation the coherent dynamics of a quasiresonance quantum system with $N{+}1$ energy levels excited  by the
monochromatic field is described by the Schr\"odinger equation written in the dimensionless form \cite{landau1, shore3, eberly shore bial1,
makarov}:
\begin{align}
 -\mathrm{i}\frac{\mathrm{d} a_{n}(t)}{\mathrm{d} t}=&f_{n+1} \exp (-\mathrm{i} \,\varepsilon_{n+1} t) a_{n+1}(t)+f_{n}
 \exp(+\mathrm{i}\,\varepsilon_{n} t) a_{n-1}(t);   \label{eq_1}\\
&n=0,1,\ldots ,N;\qquad a_{n} (t\!=\!0)=\delta_{n,0}.\nonumber
\end{align}
Here all the variables and coef{}ficients are dimensionless: $a_{n}(t)$ are the amplitudes of the probability to detect the system at the
level $E_{n}$ at the instant $t$, $\varepsilon_{n}$ is the dimensionless frequency detuning on the $n^{th}$ transition $n-1\leftrightarrow
n$. Initially the system is located at the zero level.

The field amplitude $\cal E_{\ell}$ is included into the dimensionless time and frequency detuning: $t=\Lambda \tau $,
$\varepsilon_{n}=(\omega_{n} -\omega_{\ell})/\Lambda $, where $\Lambda =\mu_{0,1} {\cal E}_{\ell} /2 \hbar$ is the Rabi frequency, and
$\omega_n =(E_{n-1}-E_n)/\hbar$ is the frequency of the $n^{\mbox{th}}$ transition in the quantum system. The parameter $N$ is any natural
number so that the number of the energy levels interacting with radiation is $N{+}1$ along with the number of equations, including
$N=\infty$.

Given the known amplitudes $a_{n}(t)$, one can determine the energy level populations $\rho_{n}(t)=\left|a_{n}(t)\right|^{2}\equiv
a_{n}(t)a_{n}^{*}(t)$. In other words, it is possible to obtain a distribution function over the energy levels at any instant. The function
completely  describes the dynamics of a quantum system.

We search the solution of (\ref{eq_1}) as one of the following expressions
\begin{equation}
a_{n}(t)=\mathrm{e}^{\mathrm{i} {\,} s_{n} {\,} t} b_{n}(t)=\mathrm{e}^{\mathrm{i} {\,} s_{n} {\,} t} \! \int_{A}^{B} \!\!\! \sigma(x)\, p_{0} \;
p_{n} (x)\; \mathrm{e}^{\mathrm{i}{\,}r{\,}x{\,} t} \; {\mathrm{d}}x,
\label{eq_2}
\end{equation}
\begin{equation}
a_{n}(t)=\mathrm{e}^{\mathrm{i}{\,} s_{n} {\,}t} b_{n}(t)=\mathrm{e}^{\mathrm{i} {\,}s_{n} {\,}t} \sum_{x=0}^{N}\sigma(x)\, p_{0} \;
p_{n} (x)\; \mathrm{e}^{\mathrm{i} {\,}r{\,}x {\,}t},
\label{eq_3}
\end{equation}
i.e. as the product of a phase factor which does not change the populations $\rho_{n}(t)=a_{n}^{*}(t)\,a_{n}(t)=b_{n}^{*}(t)\, b_{n}(t) $
by the continuous or discrete Fourier transform of  $b_{n}(t)$. Here $\omega(x)=rx$ is the continuous or discrete set of Fourier
frequencies;
\begin{equation}
S_n (x)=\sigma(x)\,p_0\;p_n (x)
\label{eq_4}
\end{equation}
is the Fourier spectrum of the time-dependent function $b_n (t)$. The spectrum is expressed in terms of an orthonormal polynomial sequence
$\left\{p_{n}(x);\right.$ $\left.n=0,1,\ldots \right\}$ with a continuous or discrete variable $x$; $\sigma(x)$ is the weight function of
the polynomials. The orthogonality relation\;is
\begin{equation}
\int_A ^B \!\!\!\sigma(x)\,p_m (x)\,p_n (x)\, {\mathrm{d}}x = \delta_{m,\,n} \quad \mbox{or} \quad \sum_{x=0}^{N}\sigma(x) p_{m} (x) p_{n} (x)=
\delta_{m,\,n} \,.
\label{eq_5}
\end{equation}
Expressions (\ref{eq_2}), (\ref{eq_3}) satisfy the initial conditions (\ref{eq_1}) because of orthogonality (\ref{eq_5}). It is obvious that the Fourier spectrum (\ref{eq_4}) of dynamical variables can be discrete (for example when $a_n (t)$, $b_n (t)$ are time-periodic
functions) or continuous. In these different cases one should make a\;choose between (\ref{eq_3}) and (\ref{eq_2}) respectively. By choosing
the relevant polynomial sequence one should take into account the number $N+1$ of energy levels and the same number of equations.

Furthermore while making the choice it is very important to take into consideration the known property of any orthonormal polynomials
\cite{suetin, preprint}:
\begin{equation}
\bar f_{n+1} \,p_{n+1} (x)+\bar f_{n} \,p_{n-1} (x)={\rm [}rx+s_{n} {\rm ]}\, p_{n} (x); \quad \bar f_{1}=1; \quad \bar f_{0}=0.
\label{eq_6}
\end{equation}
All the coef{}ficients $\bar f_{n}$, $r$, $s_{n}$ in the three-term recurrence relation (\ref{eq_6}) are given for the known orthogonal
polynomials, as well as the weight function $\sigma\left(x\right)$. The correct selection appropriate to (\ref{eq_1}) is the polynomial
sequence which satisfies the conditions $\bar f_n\equiv f_n$, i.e. the coefficients $\bar f_n$ in (\ref{eq_6}) coincide with the
coefficients $f_n$ in (\ref{eq_1}). Then (\ref{eq_2}) or (\ref{eq_3}) is the exact solution of (\ref{eq_1}), as will be shown below.

It should be particularly emphasized that the three-term recurrence relation in form (\ref{eq_6}) for an orthonormal polynomial sequence is
the most important property in solving the problem under investigation. Unfortunately the handbooks do not contain this kind of recurrence
formulas, and  both orthogonal polynomials and recurrence formulas, as a rule, are written in another standardization. But they can easily
be recast to be in form (\ref{eq_6}).

Substituting, for example, (\ref{eq_3}) into (\ref{eq_1}), we arrive at the equations
\begin{align}
\sum_{x=0}^{N}\sigma(x) p_{0} \,\mathrm{e}^{\mathrm{i} \,r\, x \,t} &\left\{f_{n+1}
\exp\bigl({-\mathrm{i}(\varepsilon_{n+1} -s_{n+1} +s_{n})t}\bigr) \,p_{n+1} (x)+\right.\nonumber\\
&\left.\;f_{n} \;\exp\bigl({+\mathrm{i}\left(\varepsilon_{n} -s_{n} +s_{n-1}\right)t}\bigr) \,p_{n-1} (x)-[rx+s_{n} ]\,p_{n} (x)\right\}=0,
\label{eq_7}
\end{align}
which are equivalent to (\ref{eq_1}) and govern the coherent dynamics of $N{+}1$-level quantum systems characterized by the parameters $N$,
$f_{n}$, $\varepsilon_{n}$. From the set of these systems we will select the systems satisfying the condition $\varepsilon_{n+1} =s_{n+1}
-s_{n}$. Then (\ref{eq_7}) is reduced to be
\begin{equation}
f_{n+1} p_{n+1}(x)+f_{n} p_{n-1}(x)=\left\{rx+s_{n}\right\}p_{n}(x),
\label{eq_8}
\end{equation}
and with the constraint $f_{n}\equiv \bar f_{n}$ (\ref{eq_8}) is valid since it coincides with recurrence relation (\ref{eq_6}) for the
orthonormal polynomials. Thus, expression (\ref{eq_3}) is the exact solution of (\ref{eq_1}), the coef{}ficients being
\begin{equation}
f_{n+1}=\bar{f}_{n+1},\qquad \varepsilon_{n+1}=s_{n+1}-s_{n}.
\label{eq_9}
\end{equation}
A similar reasoning with (\ref{eq_2}) provides the same result.

Thus, (\ref{eq_2}) or (\ref{eq_3}) is the exact solution of (\ref{eq_1}) with coef{}ficients $f_n$, $\varepsilon_n$ and the parameter $N$,
if the orthonormal polynomial sequence $\{p_n (x);\!\!\!\!\quad N\}$ in (\ref{eq_2}) and (\ref{eq_3}) contains in its recurrence formula
the coef{}ficients $\bar f_n$, $s_n$ (\ref{eq_6}) satisfying the conditions (\ref{eq_9}) linking the coef{}ficients of (\ref{eq_1}) and
(\ref{eq_6}).

It is more natural to construct spectrum (\ref{eq_4}) with the use of the equation coef{}ficients, i.e. from the characteristics of the
quantum system rather then on the basis of polynomial sequence.  However, it is more convenient to construct initially the spectrum and
then to identify coef{}ficients of (\ref{eq_1}), amplitudes $a_{n}(t)$, populations $\rho_{n}(t)$, i.e. the quantum system dynamics in the
radiation field. We set up the Fourier spectrum in a form of proper known orthonormal polynomials $p_{n}(x)$ where $x$ is a grid and
$\sigma(x)$ is a weight function. The proper polynomial sequence should be chosen on the basis of the recurrence relation and $N$. Here we
will not consider another method to construct polynomials from a recurrence formula  because this problem is dif{}ficult to obtain the
weight function. At the same time orthogonal polynomial theory is very extensive \cite{bateman2, chihara, ismail, gasper rahman, koekoek
swart}. There are a lot of various known polynomial sequences and a big field to choose a suitable sequence. It is only necessary to recast
the recurrence relations in form (\ref{eq_6}) for orthonormal polynomial sequences.

In other words, here we solve not a direct problem (to find the solution of (\ref{eq_1}) for the given quantum system characterized by $N$,
$f_n$, $\varepsilon_n$) but an inverse one: starting from the chosen polynomials, i.e. from the Fourier spectrum of the amplitudes $a_n(t)$
we construct a corresponding quantum system, determine the characteristics and build the solution for the dynamical equations (\ref{eq_1}).
At first sight it seems to be a drawback, but the major advantage of such an approach is that it allows one to classify quantum systems, to
build the unified solution not only for one system but as well for a set of ones when polynomials contain additional parameters. The theory
of orthogonal polynomials (not only classical ones) contains connections between many polynomial sequences. It allows us to transfer these
schemes to corresponding quantum systems. A considerable number of the studied polynomials will allow increasing the number of exact
solutions describing dynamics of various quantum systems in coherent fields of electromagnetic radiation.

There also exists a deeper connection between the weight function $\sigma(x)$ and corresponding quantum multilevel system because
$\sigma(x)$ determines uniquely both the polynomial sequence and the quantum system. It is safe to say that any weight function being the
''progenitor'' of polynomial sequence and corresponding quantum system defines the coherent dynamics of a quantum system in the laser
radiation field completely within the limits of the assumptions adopted here (rectangular pulse envelope, constant carrier frequency,
rotating wave approximation, and transitions between adjacent energy levels).

Theory of orthogonal polynomials is a natural mathematical formalism for the analytical description of the coherent dynamics of various
quantum systems excited by laser radiation. All the aforesaid is illustrated using some examples.

%%%%%%%%%%%%%%%%%%%
%       444       %
%%%%%%%%%%%%%%%%%%%

\section{Harmonic oscillator excited by resonance radiation}
This is the first quantum model for the description of one-dimensional movement of a particle with a mass $m$ in the parabolic potential.
The Schr\"odinger equation with the hamiltonian
\begin{equation}
{\rm \bf \hat H}_0 (z)=-\frac{\hbar}{2m} \frac{\mathrm{d}^2}{\mathrm{d} z^2}+\frac{m {\,}\omega ^2}{2} {\,}z^2, \qquad -\infty <z<\infty,
\label{eq_10}
\end{equation}
where $\omega$ is the oscillator frequency, $z$ is a dimension space coordinate, leads to the boundary-value problem ${\rm \bf \hat H}_0(z)
\varphi (z) = E \varphi (z)$ for eigenvalues and eigenfunctions \cite{landau1}:
\begin{equation}
E_n =\hbar {\,}\omega (n+\textstyle\frac{1}{2}); \quad \varphi_n (\xi)=\sqrt {\sigma (\xi)}{\,\,} {\rm \hat H}_n (\xi); \quad n=0,1,\ldots ,N=\infty.
\label{eq_11}
\end{equation}
Here $\xi =\sqrt{m{\,}\omega /\hbar}{\,\,}z$ is the dimensionless space coordinate, ${\rm \hat H}_n (\xi)$ are the orthonormal Hermite
polynomials
\begin{equation}
\int_{-\infty}^{\infty} \!\!\! \sigma(\xi)\, {\rm \hat H}_n (\xi) \;
{\rm \hat H}_m (\xi)\; {\rm d}\xi =\delta_{n,\,m},
\label{eq_12}
\end{equation}
and $\sigma (\xi) =\exp (-{\xi}^2)$ is their weight function. In contrast to the real eigenfunctions $\varphi_n (\xi)$, the total wave
functions of the harmonic oscillator stationary states have the form
\begin{equation}
\psi_n (\xi ,\tau)=\varphi_n (\xi){\,\,} \exp \biggl(-{\rm i} {\frac {E_n}{\hbar} \tau}\biggr)= \sqrt {\sigma (\xi)}{\,\,\,}
{\rm \hat H}_n (\xi) {\,\,}\exp \biggl(-{\rm i} \omega (n+\textstyle\frac{1}{2}) \tau \biggr).
\label{eq_13}
\end{equation}
They are orthonormal: $\int_{-\infty}^{\infty} \,\, \psi_n^* (\xi ,\tau) \,\,\psi_n (\xi ,\tau) \,\,{\rm d} \xi =\delta_{n,\,m}$. This is
the case of the harmonic oscillator stationary problem. %

The dynamical problem for the oscillator excited by resonance radiation with the frequency $\omega$ is described by (\ref{eq_1}), where
\begin{equation}
f_n =\sqrt n; \quad \varepsilon_n \equiv 0; \quad  N=\infty.
\label{eq_14}
\end{equation}
The equation has been solved earlier by many various methods \cite{baz2, eberly shore bial1, makarov}. To solve (\ref{eq_1}) by the method
(\ref{eq_2}), (\ref{eq_3}), one needs to choose proper orthogonal polynomials. For this purpose, it is necessary to write the recurrence
relation (\ref{eq_6}) for the orthonormal polynomial sequence with $N=\infty$. Resonance excitation of the oscillator is obvious to be
non-periodic functions $a_n (t)$ with the continuous Fourier spectra, i.e. the proper polynomials are ones of a continuous variable $x$.
Among the classical polynomials (Hermite, Laguerre, Jacobi) only the orthonormal Hermite polynomials ${\rm\hat H}_n (x), \quad
n=0,1,\ldots, N=\infty$ satisfy the required recurrence formula
\begin{equation}
\sqrt{n+1}\, {\rm\hat H}_{n+1} (x)+\sqrt{n} \,{\rm\hat H}_{n-1}(x) = \left[\sqrt 2 \,x+0\,\right] \,{\rm\hat H}_n (x); \quad n=0,1,\ldots, \infty.
\label{eq_15}
\end{equation}

Condition (\ref{eq_14}) is satisfied. The polynomial weight function is $\sigma(x)=\exp(-x^2)$, $-\infty{<}x{<}\infty$. Thus the
orthonormal Hermite polynomial sequence ${\rm\hat H}_n (x)$ provides the solution of the dynamical problem.

By substituting ${\rm\hat H}_n (x)$ into (\ref{eq_2}) and calculating the integral, one obtains the probability amplitudes
\begin{equation}
a_n (t)={\mathrm{i}}^n \frac{t^n}{\sqrt {n!}} \,\exp \,\left(-t^2 /2\right),
\label{eq_16}
\end{equation}
and then the distribution function $\rho_n (t)=a_{n}(t)\,a_{n}^{*}(t)$ (Poisson distribution), i.e. the energy level populations are
\begin{equation}
\rho_n (t)=\frac{1}{n!} \,{<\!\!n\!\!>}^n \,{\mathrm{e}}^{-<n>}; \quad {<\!\!n\!\!>}=t^2 ; \quad n=0,1,\ldots,\infty.
\label{eq_17}
\end{equation}
The Poisson distribution parameter $<\!\!n\!\!>$ is the average number of quanta absorbed by the oscillator at the instant $t$ because
\begin{equation}
\sum_{n=0}^{\infty} n \rho_n =<\!\!n\!\!>,
\label{eq_18}
\end{equation}
and the average energy absorbed is
\begin{equation}
<\!E_n\!> \,=\hbar \,\omega \sum_{n=0}^{\infty} n \,\rho_n =\hbar \,\omega <\!\!n\!\!>.
\label{eq_19}
\end{equation}
 Although this is the known result, the example, nevertheless, shows clearly that recurrence relation (\ref{eq_6}) for orthonormal
 polynomials plays a great role in the choice  of a suitable polynomial sequence.

%%%%%%%%%%%%%%%%%%%
%       555       %
%%%%%%%%%%%%%%%%%%%

\section{Harmonic oscillator excited by non-resonance\\ radiation}
In this case the oscillator dynamics is governed by (\ref{eq_1}) with $f_n$, $N=\infty$ as well but the frequency detunings $\varepsilon_n
\equiv \varepsilon \ne 0$ are not equal to zero on the transitions. The oscillator is out of resonance; excitation is known to be limited,
i.e. $a_n (t)$ are periodic functions of time with the discrete Fourier spectra (\ref{eq_4}), more exactly their spectra are functions of a
discrete variable\;$x$. Therefore the solution of the problem should be sought in form (\ref{eq_3}) and proper polynomials have to satisfy
(\ref{eq_6}) with
\begin{equation}
\bar f_n =\sqrt n ,\quad s_n = \varepsilon n +\mbox {const}, \quad N=\infty.
\label{eq_20}
\end{equation}
Among the classical polynomials of a discrete variable, the orthonormal Charlier polynomial sequence \cite{nikif susl uvar}
\begin{equation}
{\rm\hat C}_{n}^{(\mu)}(x);\!\!\!\quad x=0,1,\ldots,\infty ;\!\!\!\quad n=0,1,\ldots,\infty ;\!\!\!\quad \sigma (x)=\frac{\mu ^x}{x!}
e^{-\mu}; \!\!\!\quad \mu >0
\label{eq_21}
\end{equation}
satisfies (\ref{eq_6}) with
\begin{equation}
\bar f_n =\sqrt n ,\quad r=-1/\!\sqrt \mu ;\quad s_n = \mu^{-1/2}\,n+\mu^{1/2}.
\label{eq_22}
\end{equation}
Then series (\ref{eq_3}) with $p_n (x)={\rm\hat C}_{n}^{(\mu)}(x)$ gives the exact solution of (\ref{eq_1}) where $f_n =\sqrt n$;
$\varepsilon_n \equiv \varepsilon =\mu^{-1/2}$. Owing to the parameter $\mu$, the orthonormal polynomial sequence gives the solution
describing excitation by radiation with an arbitrary frequency.

The solution reads
\begin{equation}
a_n^+ =\exp \biggl({\rm i} \left(n\varepsilon +\frac{1}{\varepsilon}\right)t \biggr)\,\,\frac{1}{\sqrt {n!}} \,\,\biggl(\frac{\theta}{\varepsilon}\biggr)^n
\exp \biggl(-\frac{\theta}{\varepsilon ^2} \biggr),\quad \theta = 1-\exp ({-\rm i} \,\varepsilon \,t ),
\label{eq_23}
\end{equation}
for $\varepsilon>0$. For $\varepsilon<0$ the solution is $a_n^- (t)=(-1)^n a_n ^+ (t)$. The distribution function (energy level
populations) has the form
\begin{equation}
\rho_n (t)=\frac{1}{n!} \,{<\!\!n\!\!>}^n \,{\mathrm{e}}^{-<n>}; \quad {<\!\!n\!\!>}=\biggl(\frac{2}{\varepsilon}\biggr)^2
\sin^2\biggl(\frac{\varepsilon}{2} \,\biggr); \quad n=0,1,\ldots,\infty.
\label{eq_24}
\end{equation}
This is the Poisson distribution as with the preceding example (\ref{eq_17}). Now the distribution parameter ${<\!n(\varepsilon , t)\!\!>}$
is periodically time-dependent.

In the limit $\varepsilon \to \!0$ (resonance excitation) we obtain  $<\!\!n\!\!>_{\mbox{\scriptsize res}}\,=t^2$ as in (\ref{eq_17}). Thus
the solution (\ref{eq_24}) obtained with the use of the Charlier polynomials contains both non-resonance and resonance cases.

The availability of additional parameters in a polynomial sequence is a favorable property since it allows us to obtain the solutions for
various conditions of excitation as we have just seen. In this example the parameter $\mu$ is contained in $s_n$ of (\ref{eq_6}). If a
parameter is contained in $\bar f_n$ of the recurrence relation we obtain the solution not only for unique quantum system but for some
family of systems as well, as is shown below.

%%%%%%%%%%%%%%%%%%%
%       666       %
%%%%%%%%%%%%%%%%%%%

\section{Kravchuk polynomials and Kravchuk oscillators}
Now we turn to the Kravchuk polynomial sequence ${\rm K}_{\,n}^{\,(p)} (x;N)$; $n=0,1,\dots,N$ of a discrete variable $x=0,1,\dots,N$. The
sequence contains two parameters $0<p<1$ and any preassigned positive integer number $N$. The polynomials are orthogonal with respect to
the weight function
\begin{equation}
 \sigma(x)=\biggl(\!{N\atop {x}} \!\biggr) p^{\,x} q^{N-x};\quad \!\!\!x=0,1,\ldots,N;\quad
\!\!\!\! \biggl( \! {N\atop {x}} \!\biggr)=\frac{N!}{x!\left(N-x\right)!};\quad \!\!0<p<1;\quad \!\!p+q=1;
\label{eq_25}
\end{equation}
In statistical terms it is called the binomial distribution \cite{jonson kotz kemp}. It was M.\,Kravchuk who used it for the first time to
construct the orthogonal polynomial sequence, these polynomials bear his name \cite{nikif susl uvar, bateman2}. The Kravchuk polynomials
generalize the Hermite ones. Using the known recurrence relation for nonnormalized ${\rm K}_{\,n} ^{\,(p)} (x;N)$ and their norms $d_n$, it
is easy to write (\ref{eq_6}) for the orthonormal Kravchuk polynomials $p_n (x)={\rm K}_{\,n}^{\,(p)} (x;N) / d_n$. The recurrence relation
has coef{}ficients \cite{preprint}
\begin{equation}
\bar f_{n} =\left[n(N\!+\!1\!-\!n)/N\right]^{1/2} ;\quad r=1/\sqrt{p\,qN};\quad s_{n} = r(p-q)n-rpN.
\label{eq_26}
\end{equation}
This means that (\ref{eq_3}) is the exact solution of (\ref{eq_1}) with the coef{}ficients
\begin{equation}
f_{n} =\sqrt{\frac{n(N\!+\!1\!-\!n)}{N}} ;\quad \!\!\varepsilon_{n}=s_n -s_{n-1} \equiv \varepsilon =r(p-q)=(p-q)/\sqrt{p\,qN}.
\label{eq_27}
\end{equation}
From $f_n$ one can see that (\ref{eq_3}) describes the dynamics of the one-parameter family of $(N{+}1)\mbox{-}$level quantum systems with
the dipole moment function $f_n$. The family contains various systems including the two-level system (with $N=1$; $f_1 =1$; $f_2 =0$) and
the harmonic oscillator (in the limit $N \to \infty$; $f_n =\sqrt n$). Thus, both these fundamental quantum models are the special cases of
the Kravchuk oscillators. Since the frequency detuning $\varepsilon_{n} \equiv \varepsilon$ does not depend on the transition number, the
energy levels of the Kravchuk oscillator are arranged equidistantly. Each $N$-oscillator is excited by monochromatic radiation with an
arbitrary frequency, whereas the detuning $\varepsilon$ is defined by parameters $N$ and $p$ which can be varied. The value $p=1/2$
corresponds to resonant excitation. The dipole moment function $f_n$ (for $N<\infty$) increases and then decreases with $n$, the function
is symmetric: $f_n=f_{N+1-n}$, $n=0,1,\ldots,N$. The Kravchuk oscillator with $N=2$ is the tree-level equidistant equal-Rabi model ($f_1
=f_2 =1$, $f_3 =0$.) For the other $N$ we have various equidistant-level systems with specific energy dependence $f_n$ (\ref{eq_27}) of
dipole transitions matrix elements.

%%%%%%%%%%%%%%%%%%%
%       777       %
%%%%%%%%%%%%%%%%%%%

\section{Coherent dynamics of Kravchuk oscillators family}

So the exact solution is defined by the sum (\ref{eq_3}) with the normalized Kravchuk polynomials $p_{\,n}(x)={\rm \hat
K}_{\,n}^{\,(p)}(x;N)$. It can be calculated using the representation of the Kravchuk polynomials through hypergeometric function
\cite{bateman2}:
\begin{equation}
{\rm K}_{\,n}^{\,(p)} (x;N)=(-1)^{n}
\biggl(\!{N\atop x}\!\biggr)p^{\,n}\,\,_{2}F_{1} \left(-n,-x;-N;1 /p\right).
\label{eq_28}
\end{equation}
The exact solution is thus found to be
\begin{equation}
 a_{n}(t)=\left[\biggl(\!{N\atop x}\!\biggr)(p\,q)^{n} \right]^{1/\,2}\!\!\left(p\,\mathrm{e}^{{\mathrm{i}}\,
r\,t} +q\right)^{N-n} \left(\mathrm{e}^{\mathrm{i} \,r \,t} -1\right)^{n}\,\,\exp \{\mathrm{i}\,[(p-q)n-pN\,]\,r\,t\}.
\label{eq_29}
\end{equation}
Energy levels populations are expressed as a binomial distribution
\begin{equation}
 \rho_{n}(t)=\biggl(\!{N\atop n}\!\biggr)
\,  \left[1-y(t)\right]^{N-n} y^{n}(t),\qquad\;
y(t)=\frac{1}{N} \left(\frac{2}{r}\right)^{2} \sin^{2} \left(\frac{r}{2} \,t\right),\quad n=0,1,\ldots,N.
\label{eq_30}
\end{equation}

\begin{figure}[tbhp]
\centering
\includegraphics[width=0.5\textwidth]{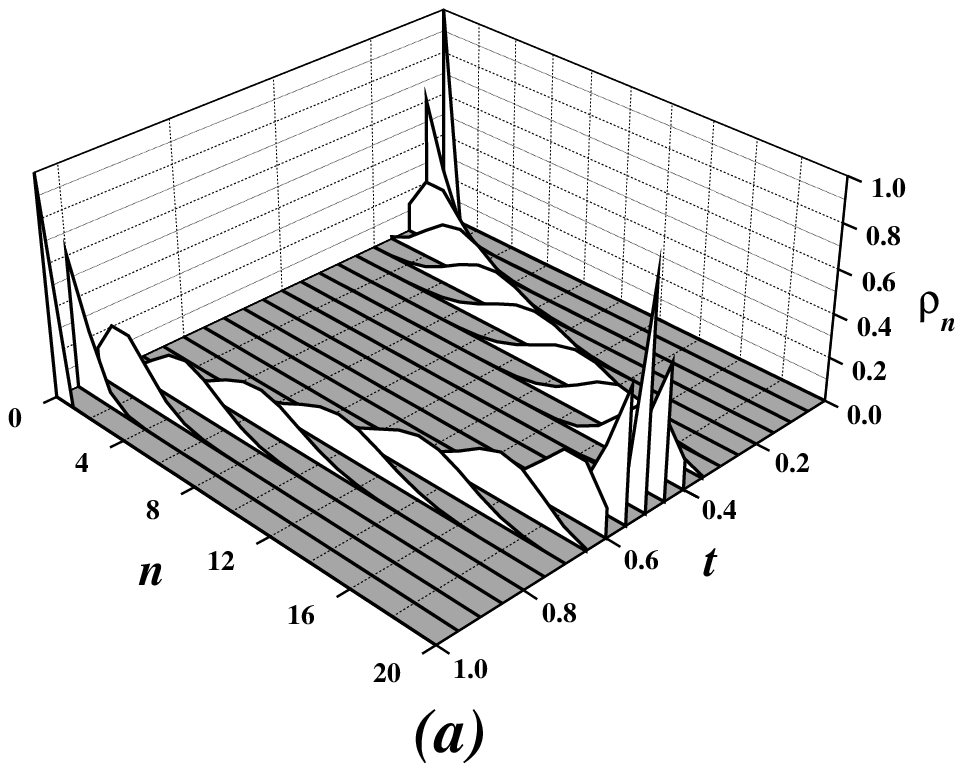}\includegraphics[width=0.5\textwidth]{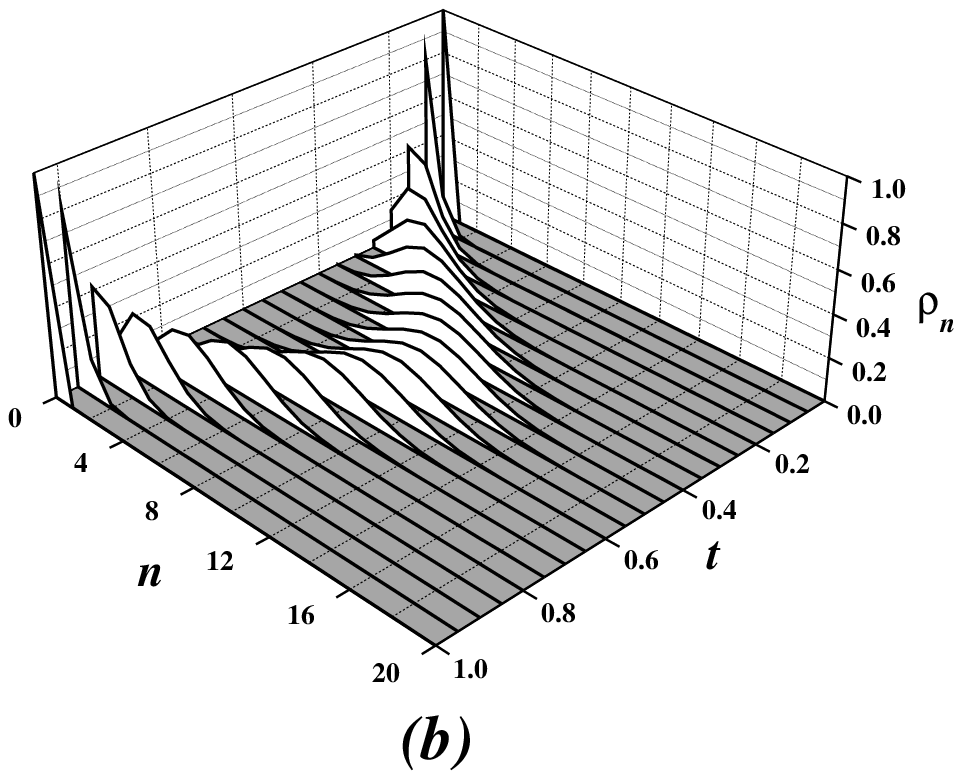}
\caption{Kravchuk oscillator: excitation dynamics of the system with finite number of levels ($N=20$); figure 2(a) resonant case
($\varepsilon=0$), figure 2(b) nonresonant case ($\varepsilon=0.5$).}
\end{figure}
The distribution parameter $y(t)$ depends on the proper time $rt$. The solution can be expressed not only through  $N$ and $r$ but directly
through the parameters of (\ref{eq_1}) $N$ and $\varepsilon$ as well, because
\begin{equation}
r=1/\sqrt{p\,qN} =\sqrt{\frac{4}{N} +\varepsilon^{2}} \,.
\label{eq_31}
\end{equation}
Excitation dynamics of the Kravchuk oscillator with finite number $N+1=21$ of energy levels is shown in Figure 2 both in and beyond
resonance. The binomial distribution of the level populations exists at any instant. The movement is periodic: the oscillator stores energy
from radiation and then restores it to the field. Outside of resonance only a part of levels are populated during the interaction with
radiation.

%%%%%%%%%%%%%%%%%%%
%       888       %
%%%%%%%%%%%%%%%%%%%

\section{Special cases of the Kravchuk oscillators and their coherent dynamics}
In the limit $N\to \infty$, $y(t)\to 0$ so that $\lim_{N\to\infty}N y(t)=\lambda(t)$, the two-parametric binomial distribution
(\ref{eq_30}) is transformed to the Poisson distribution \cite{jonson kotz kemp}
\begin{equation}
\rho_{n}(t)=\frac{1}{n!}\, \lambda^{n}(t) \, \mathrm{e}^{-\lambda(t)} ;\quad \lambda(t)=\left(\frac{2}{\varepsilon} \right)^{2} \sin^{2}
\left(\frac{\varepsilon}{2} \,t\right);\quad n=0,1,\ldots,\infty.
\label{eq_32}
\end{equation}
Here $\lambda(t)$ is a distribution parameter. In this case $f_{n}=\sqrt n$, i.e. we obtain the solution (\ref{eq_24}) for coherent
excitation of the harmonic oscillator by radiation with the frequency detuning $\varepsilon$. In the resonance condition ($\varepsilon=0$),
$\lambda(t)=t^{2}$ and the solution takes the form (\ref{eq_17}).

Another borderline case $N{=}1$ of the solution (\ref{eq_30}) corresponds to coherent excitation of the  two-level model. In this case
$f_{1}=1$ (a single transition) and one obtains
\begin{equation}
 \rho_n (t)=\left(1-y\right)^{1-n} y^{\,n};\quad\;
y\equiv y(t)=\frac{1\,}{\Omega^{\,2}} \sin^2\left(\Omega \,t\right);\quad \Omega =
\sqrt{1+\left(\varepsilon\left/2\right.\right)^2} ; \quad n=0,1.
\label{eq_33}
\end{equation}
It coincides with the known solution \cite{landau1}. The levels populations are $\rho_{0}(t)=1-y,\quad \rho_{1}(t)=y$. At resonance
excitation $\Omega =1$, $y=\sin^{2}t$, levels populations are $\rho_0=\cos^2t$, $\rho_1=\sin^2t$.

%%%%%%%%%%%%%%%%%%%
%       999       %
%%%%%%%%%%%%%%%%%%%

\section{Conclusions}

Starting from the phenomenological model of a quasiresonance medium excited by coherent radiation, a method is proposed to construct exact
analytical solutions of the equations for the probability amplitudes. The method uses continuous or discrete Fourier transforms of the
amplitudes, where Fourier spectra are expressed in terms of some orthonormal polynomial sequence multiplied by its weight function. Some
exact solutions are obtained and the distribution functions over the quantum system energy levels depending on time and on frequency
detuning are presented. The distributions follow from Schr\"odinger equation exact solutions and give the complete dynamical description of
laser-excited  quantum multilevel systems when any relaxation processes are eliminated.

The Kravchuk oscillator family as an integrable model has been constructed to describe coherent excitation dynamics of multilevel resonance
media. The model is based on the use of the Kravchuk orthogonal polynomials. Kravchuk oscillator excitation dynamics is described with the
binomial distribution of energy level populations and a distribution parameter depends on excitation conditions. Two  basic models known in
quantum physics -- the harmonic oscillator and the two-level system are special representatives of the Kravchuk oscillator family.

Physical interpretation of the method is expounded. The Fourier spectra of the amplitudes are expressed in terms of otrhonormal polynomials
of a continuous or discrete variable which has meaning of a dimensionless frequency. There is the  one-to-one correspondence between the
mathematical structures (orthonormal polynomials, their weight function with its range of definition) used and quantum system
characteristics (energy levels, dipole moment matrix elements, transitions frequency detunings) along with the dynamical equation
coef{}ficients. The recurrence relation for orthonormal polynomials is shown to play the keynote role in the selection of a suitable
polynomial sequence to construct the exact solutions for the coherent dynamics of various quantum systems.

In this problem a weight function is shown to be conceptually the only generative object because it defines an orthogonal polynomial
sequence, the recurrence formula, quantum system characteristics, dynamical equation coef{}ficients, Fourier spectra of the probability
amplitudes, the probability amplitudes proper, energy level populations, and coherent dynamics of the relevant quantum system in the end.
If the weight function and its polynomial sequence, respectively, contain some parameters, the solution describes the dynamics of the
quantum systems family excited under various conditions.

Orthogonal polynomials are known to be used in stationary quantum problems as well to obtain  eigenvalues and eigenfunctions of quantum
oscillators \cite{landau1}, including some polynomials of a discrete variable \cite{atakishiev suslov}. In the latter case physical
interpretation of the results is adaptable to discrete physical space, to discrete quantum mechanics \cite{odake sasaki}. As it is shown
above, in the traditional orthodox quantum mechanics the arguments of orthogonal polynomials differ in physical nature for stationary and
dynamical problems: in the former case the argument is a dimensionless space coordinate but in the latter case the argument of the
polynomials is the Fourier-frequency of time-dependent functions, i.e. of the probability amplitudes of a quantum system. In a steady-state
problem the polynomials are defined in a domain of physical space but in the non-stationary one they are defined in the Fourier space that
can be both continuous and discrete. The last case is implemented when the probability amplitudes are periodic functions of time. This case
is achieved  if radiation interacts with a finite number of energy levels, that is radiation induces a finite number of transitions in a
quantum system. Such is indeed the case of practical interest. Therefore orthogonal polynomials of a discrete variable are the most useful
tool to solve the problems on coherent excitation of multilevel systems in the common quantum mechanics.

There are a lot of various orthogonal polynomial sequences which can be used to construct exact solutions for  dynamics of diverse quantum
multilevel models of the laser excited quasiresonance media.

%%%%%%%%%%%%%%%%%%%
%      101010     %
%%%%%%%%%%%%%%%%%%%

\end{document}